# Tailoring energy barriers of Bloch-point-mediated transitions between topological spin textures

Yu Li[1,2*], Yuzhe Zang[1], Runze Chen[1], Christoforos Moutafis[1*]

[1]Nano Engineering and Spintronic Technologies (NEST) Group, Department of Computer Science, University of Manchester, Manchester M13 9PL, United Kingdom.

[2]Frontier Institute of Chip and System, Fudan University, Shanghai, 200438, China.

*Email: YuLi.nano@outlook.com; Christoforos.Moutafis@manchester.ac.uk.

## Abstract

Magnetic skyrmions are nanoscale spin textures that their thermal stability originates from the nontrivial topology in nature. Recently, a plethora of topological spin textures have been theoretically predicted or experimentally observed, enriching the diversity of the skyrmionic family. In this work, we theoretically demonstrate the stabilities of various topological spin textures against homochiral states in chiral magnets, including chiral bobbers, dipole strings, and skyrmion tubes. They can be effectively classified by the associated topological Hall signals. Multiple transition paths are found among these textures, mediated by Bloch-point singularities, and the topological protection property here can be manifested by a finite energy barrier with the saddle point corresponding to the Bloch-point creation/destruction. By carefully modulating the local property of a surface, such as interfacial DMI induced by breaking the structural symmetry, the energy landscape of a magnetic system can be tailored decisively. Significantly, the proposed scenario also enables the manipulation of stabilities and transition barriers of these textures, even accompanied by the discovery of ground-state chiral bobbers. This study may raise great expectations on the coexistence of topological spin textures as spintronics-based information carriers for future applications.

### I. Introduction

Chiral magnets can host a rich variety of spin textures due to their inherent broken inversion symmetry of crystal structures[1–9], including many Ge-[1–5] and Si-[6–8] based B20 compounds, and some insulating magnets such as $Cu_2OseO_3$[9]. Their stabilisation highly relies on the occurrence of the Dzyaloshinskii-Moriya interaction (DMI)[10–12]. Magnetic skyrmions are usually referred to as two-dimensional particle-like spin configurations[13], and their nontrivial topology can be characterised by the topological charge[14]:

$$Q_{2D} = \frac{1}{\phi_0} \int B_z^e \, d^2r = \frac{1}{4\pi} \int \bm{m} \cdot [\partial_x \bm{m} \times \partial_y \bm{m}] \, dxdy, \qquad (1)$$

which is the effective flux of the emergent magnetic field $B_z^e = \frac{\hbar}{2}\mathbf{m} \cdot [\partial_x\mathbf{m} \times \partial_y\mathbf{m}]$ originating from the spatial variation of the magnetisation field **m**, with $\phi_0$ indicating the flux quantum $2\pi\hbar/q^e$ ($q^e = \pm 1/2$ for minority/majority spins)[15]. For a single skyrmion, the $Q_{2D}$ corresponds to an integer value $\pm 1$ with the sign determined by the skyrmion polarity. However, in bulk chiral magnets with a considerable size along the thickness direction, more diverse and complex spin textures exist (Fig. 1d–h). Single skyrmions are stabilised as a form of skyrmion tubes through the samples[1,2,16–18] (Fig. 1d). In addition, Bloch points (BPs) are point-like topological defects where the magnetisation vanishes in the centre and play crucial roles in mediating skyrmion annihilation processes[2,14,18–20]. Chiral bobbers are formed by skyrmion tubes partially penetrating into a sample with a finite depth and ending with Bloch points (Fig. 1f–h), whose thermal stability is attributed to the geometrical confinement of surfaces[4,20–22]. The vicinity of Bloch points at both ends can lead to spin textures referred to as dipole strings[23–26] (Fig. 1e). These spin textures involve three-dimensional spin swirling in nontrivial manners, and usually possess complex distributions of the emergent magnetic field $B_z^e$. The collective effect of $B_z^e$ is associated with unique topological properties for different textures, and the topological Hall effect enables a concise solution for detecting and distinguishing these structures using electronic transport measurements[8,22,30–35]. Most of the existing studies focused on the correlation of topological Hall signal and the topological charge in quasi-two-dimensional systems[32–34]. However, concerning the above-mentioned spin textures, such transport properties should also take into account the depth-dependent complexity. Redies et al. analysed the unambiguous signals of chiral bobbers and skyrmion tubes, which are relevant to the volume that a spin texture occupies in the sample[22]. Our previous work also revealed the topological Hall response of skyrmion "chains" in magnetic multilayers, and its stepwise variation is induced by the Bloch-point hopping through the multilayer[35]. It can be speculated that Bloch points play key roles in modulating the topological behaviour in an explicit form of transport property. Furthermore, before further experimental implementation in actual spintronic devices, studies on thermal stabilities and transitions among these textures are necessary and have drawn much attention[6,36–41]. The energy barriers of transitions could imply the robustness of specific textures against external excitations, especially the thermal fluctuations. These properties can be explored by investigating the minimum energy paths between two equilibrium states. At the same time, the topology of skyrmions endows them with stability[6,39,42], which suggests that topological constraints prohibit continuous deformations into other textures with different topological charges. For example, the topological charge of a skyrmion keeps relatively invariant ($Q_{2D} = \pm 1$) under its expansion or reduction in size, whereas a considerable energy barrier exists when it unwinds into a topologically trivial ferromagnetic state (also called homochiral state in the following sections, $Q_{2D} = 0$). These topological charge descriptions essentially correspond to the two-dimensional winding number in either an infinite plane $R^2$ or a unit sphere $S^2$ around a point[14], and is inappropriate to be directly migrated to a three-dimensional regime. It thus raises an intriguing question of how the topological protection property assists topological spin textures with complex magnetization profiles to maintain the topological charge in a depth-dependent manner.

In this work, we present that various topological spin textures, including chiral bobbers, dipole strings, and skyrmion tubes, can be stabilised in chiral magnetic nanodisks. Their associated topological Hall responses are analysed, which points to distinct magnitudes of signals regarding the type of textures. Then we calculate several possible transition paths between the HC and SkT states, which reveals the limitation of the topological protection property within finite magnetic systems. A scenario for tailoring the thermostabilities as well as energy barriers of transitions is also proposed via local modulation of surface properties, such as additionally breaking the structural symmetries. It can be regarded as a superior solution for switching between the topological spin textures apart from conventional magnetic field or thermally induced mechanisms.

## Results
**Stabilisation and associated topological signature.**

The phase diagram of magnetic textures in bulk chiral magnets has been extensively studied[20,43,44]. At zero (or small) external magnetic field, the magnetisation prefers an in-plane alignment due to the weak magnetocrystalline anisotropy in pristine crystals such as FeGe[45], corresponding to a helical state (Fig. 1a); when a strong out-of-plane field is applied, a conical (Fig. 1b) or even a saturated ferromagnetic state (Fig. 1c) becomes energetically favourable. The magnetisation profiles of these three states periodically propagate by the length of $L_D$. Regarding their homogeneous spin along the propagation direction (wave vector), we refer to them as homochiral (HC) states in order to distinguish them from the topological nontrivial spin textures which will be discussed in the following. Magnetic skyrmions in three dimensions are stabilised in the form of skyrmion tubes (Fig. 1d). Although the ideal solutions of skyrmions modulated by the bulk DMI are Bloch-type, the lack of neighbouring spins near the top and bottom surfaces induces distortions of the chiralities along a skyrmion tube[16,17,43,46], i.e., the chirality of the skyrmion profile is $\pi/2$ (pure Bloch-type) in the middle plane, but gradually increases/decreases with the mixture of a Néel-type profile when moving towards the top/bottom surface. As a result, this "surface twist" effect will induce a spatial inhomogeneity of energy distributions throughout the film, especially the magnetostatic energy[18]. Moreover, dipole strings (DSs) are a type of topological spin textures located in the interior of the system without the occupation of surface states (Fig. 1e), accompanied by Bloch-point (BP) pairs holding opposite topological charges at the two ends[25]. In contrast, the surface confinement facilitates the stabilisation of chiral bobbers (CBs) with the vicinity of single Bloch points at the end[4,20]. If the sample thickness is beyond a critical value (around half of $L_D$ [20]), chiral bobbers can exist as local states near the top (Fig. 1f) or bottom (Fig. 1g) surface. Pairs of chiral bobbers (CB pairs) may also be stabilised at both sides (Fig. 1h) where the couplings between the two Bloch points can be relatively weak.

The total energies of the aforementioned topological spin textures are illustrated in Fig. 1i as a function of the external magnetic field $B_{\text{ext}}$. They are plotted as the relative magnitude with reference to that of the HC state (black circles). Specifically, in the phase diagram $B_{\text{ext}}$ can be expressed in the units of the critical field $B_0 = \mathcal{D}^2/2\mathcal{A}M_s$ ($\sim 0.37\,\text{T}$ for FeGe)[43,47], which indicates the transition field between the conical and saturated states in an ideal bulk system. However, this critical-field consideration in literature does not include the contribution of magnetostatic interactions, which we find plays essential roles in finite systems, such as nanodisks. Therefore, in this work, we denote an adjusted critical field $B_0' = 0.57\,\text{T}$, extracted from the numerical results of conical-saturated transitions (see Supplementary note 1 for more details). The ground states of the nanodisk system are dependent on the external field: the global energy minimum is the HC state in a wide range; as the stability of in-plane helical modes is weakened by $B_{\text{ext}}$, the out-of-plane modulation starts to play a role with the emergence of metastable stacked spin spirals (a mixture of the helical and conical states)[43], and thus yields a ground-state skyrmion tube (in the range of $0.39B_0' - 0.84B_0'$).

All the other topological spin textures, including dipole strings and chiral bobbers are metastable in a wide range of the external field $B_{ext}$. The stabilisation of single dipole strings in bulk chiral magnetic crystals can be assisted by the coupling to adjacent spin textures, such as skyrmion tubes[25]. In our case, their stabilities are enhanced by the top and bottom surfaces, and the mechanism is similar to that of chiral bobbers where a sufficiently large energy is required to push the Bloch points out of the surfaces. Different annihilation fields (dashed lines in Fig. 1i) of dipole strings and chiral bobbers could indirectly reflect different barrier heights of their Bloch-point-mediated transitions. The application of a magnetic field $B_{ext}$ (along $+z$ direction) pushes Bloch points with positive/negative charges (BP$^+$/BP$^-$) to move downwards/upwards. As the magnitude of $B_{ext}$ increases, the annihilation of a dipole string is initiated by the mutual combination and annihilation of two opposite-charge Bloch points (dashed green lines in Fig. 1i)[25]. In contrast, the annihilation process of a chiral bobber (or chiral-bobber pairs) occurs with the Bloch point(s) being pushed out of surface(s). Although Bloch points are commonly simplified as magnetic monopoles, it should be noted that their interaction with $B_{ext}$ field cannot be completely analogous to the motion of electric charges under an external electric field, because the topological charge of a Bloch point is the dot product of its polarity and vorticity[48,49] [Eq. (1)]. They are relevant to the spin configuration parallel and perpendicular to the field direction, respectively, and only the former term determines the propagation direction[18].

Furthermore, Müller et al. demonstrated that a dipole string can be elongated to a length in several orders of $L_D$[25]. In this work, we observe similar characteristics in single chiral bobbers, whose penetration depth is also modulated by $L_D$ (cyan/blue triangles in Fig. 1i). The existence of finite energy barriers even enables a "hysteretic switching" operation within single chiral bobbers (shaded cyan/blue region), i.e., a "short" chiral bobber (penetration depth $L_P < L_D$) can be switched to an elongated bobber ($L_P > L_D$) by altering the external field lower than $0.70B_0'$ (0.40 T), and the state can be recovered by simply raising the field higher than $0.98B_0'$ (0.56 T). Supplementary note 2 also shows that even more types of chiral bobbers can be stabilised in a thicker nanodisk ($L = 4L_D$).

The discovery of the above-mentioned (meta)stable topological spin textures could significantly enrich the diversity of skyrmion-like nontrivial quasiparticles. Their electronic-transport fingerprints, which stem from the emergent magnetic properties, are highly relevant to technologies such as erecting distinct states, which is paramount for the implementation of skyrmionic devices[22]. Previous numerical and experimental works have shown that the topological Hall (TH) effect of a spin texture is correlated to its three-dimensional extent[22,34,35]. In the following, we further identify the topological Hall responses arising from these complex three-dimensional textures. Our demonstrations are based on the calculation performed on defect-free nanodisks that are attached to four terminals (schematic are shown in Fig. 2a), where the spin textures are placed in the centre. We also assume an adiabatic regime with a strong coupling $J_\text{H} = 5t$, thus the moving electrons will have their spin aligned to the direction of the local magnetisation from the textures[31,35]. Fig. 2b illustrates the topological Hall resistivities $\rho_{xy}^\text{TH}$ as a function of the Fermi energy $E_\text{F}$, whose magnitudes are substantially distinguishable between different textures. General behaviours of the spin injection here are in qualitative agreement with those in two-dimensional systems shown in previous work[30–32]. When $E_\text{F}$ increases from zero and starts to overlap with the energy bands, spin injection and deflection are initiated with the presence of finite $\rho_{xy}^\text{TH}$. Its variation, especially the sign change from positive to negative, is explained in Ref.[31], where the carrier compositions in the transport channels are dominated from the electron type to hole type. Due to the finite size of the nanodisks, the calculations are carried out with a limited number of modes[32], hence $\rho_{xy}^\text{TH}$ fluctuations cannot be completely prevented, especially near the band edges[31].

Although the magnitude of $\rho_{xy}^\text{TH}$ signal is dependent on the $E_\text{F}$ and the type of spin textures, the intrinsic connection between $\rho_{xy}^\text{TH}$ and $B_z^\text{e}$ still exists. We first normalise the $\rho_{xy}^\text{TH}$ by rescaling the values regarding the two extreme states: HC state (black) and SkT (red) at each Fermi energy $E_\text{F}$. As the $\rho_{xy}^\text{TH}$ manifests the collective effect of the emergent magnetic field $B_z^\text{e}$ radiated by the whole three-dimensional system, instead of referring to the general expression of two-dimensional topological charge (Eq. (1)), we introduce a "global topological charge", defined as the volume integral of the emergent magnetic field $B_z^\text{e}$ over the whole three-dimensional magnetic system:

$$Q_\text{g} = \frac{1}{\phi_0} \int_V B_z^\text{e}\, \text{d}^3 r. \tag{2}$$

As a result, the normalised $\rho_{xy}^\text{TH}$ of each spin texture is approximately invariant in a wide range of $E_\text{F}$, and the magnitude is further proportional to $Q_\text{g}$ (Fig. 2c). It thus ensures the sensitivity of classifications of topological spin textures regarding the topological Hall signals via electronic transport measurements. Moreover, it should be noted that the topological Hall resistivities $\rho_{xy}^{\text{TH}}$ of some spin textures have relatively similar magnitudes, e.g., dipole strings with short chiral bobbers, and long chiral bobbers with chiral-bobber pairs. The existence of monopole-like Bloch points significantly enhances the $B_z^e$ field in local regions comparing to the $B_z^e$ of skyrmion tubes[22], thus the $\rho_{xy}^{\text{TH}}$ can be modified by the propagation of Bloch points through the film, which will be discussed in the following section.

**Minimum energy path mediated by Bloch points.**

We have shown that various topological spin textures can be stabilised in chiral magnets. To identify the stabilities of metastable states against small excitations such as thermal fluctuations, we calculate the minimum energy paths of the transitions among these textures. Due to the complexity of the phase space in the three-dimensional magnetisation profile, it will be unrealistic in this study to consider all possible paths. Regarding the magnetic configurations of these textures and inspired by their evolutions under the magnetic field[18,25], we focus on three transition paths between the equilibrated homochiral state and a single skyrmion tube, via the intermediate presence of textures shown in Fig. 3a. A type-I transition path is given by the formation and extension of a single chiral bobber from the top/bottom surface, until the chiral bobber reaches the other surface, while this process is also accompanied by the Bloch point being eliminated. A type-II transition path can then be characterised by a pair of chiral bobbers at both surfaces and ends up with the mutual annihilation of two Bloch points inside the system. A type-III transition path starts with a dipole string with two Bloch points, and the skyrmion tube forms by the extension of the dipole string. As we consider sufficiently large nanodisks with the diameter of $4L_D$, the paths mediated by lateral boundaries are neglected[39].

The searching of the minimum energy paths is implemented by the geodesic nudged elastic band (GNEB) method. Before the calculation of a transition path, the initial guess of the band is set up by pre-defining several local energy minima (including single CB, CB pair, and DS), and these minimum energy states are connected with the HC state and SkT through multiple images calculated by linear interpolation methods, so the guessed transition follows a spatially homogeneous manner with all the magnetisation vectors flipped simultaneously. Then the band is relaxed by the effective forces $\mathbf{F}_i$ applied on each image $\mathbf{Y}_i$, composed of (1) the component of the energy gradient perpendicular to the band $\mathbf{F}(\mathbf{Y}_i)|_\perp$ as well as (2) a spring force along the band $\mathbf{F}(\mathbf{Y}_i)|_\parallel$. The spring force guarantees the images to be evenly distributed rather than overlapped or fallen into the energy minima all together.

Fig. 3b–d illustrate the minimum energy paths (black rectangles), associated with the variations of Bloch-point position (red triangles), global topological charge $Q_g$ (blue crosses) and topological Hall resistivity $\rho_{xy}^{\mathrm{TH}}$ (purple diamonds). Surprisingly, although the initial bands are calculated by the linear interpolation method, all the calculations converge to the paths mediated by Bloch-point dynamics. Between each two local energy minima along the paths, we find the presence of saddle points are always accompanied by the creation or destruction of Bloch points(s) (dashed lines in Fig. 3b–d). These phenomena during the transitions should be in correspondence with the topological protection properties[6,39,42]. Due to the discrete nature of magnetic systems, this property is not infinitely robust, but in practice, is manifested by finite energy barrier[14], which has been well captured by the numerical discretisation of the micromagnetic model[18,19,50]. Based on our calculations, the energy barriers of the Bloch-point creation (destruction) are dependent on the types of the original topological spin textures. For example, the energy barrier for creating a Bloch point from the HC state to tCB/bCB is $\sim 7.9\,\mathrm{eV}$, lower than the barrier from the SkT to tCB/bCB ($\sim 13.8\,\mathrm{eV}$). Importantly, the formation of a dipole string needs to climb over a twice higher energy barrier ($\sim 15.9\,\mathrm{eV}$) than creating a chiral bobber ($\sim 7.9\,\mathrm{eV}$), which may be attributed to two Bloch points created during the process. In contrast, for a single dipole string only a tiny barrier ($\sim 0.5\,\mathrm{eV}$) exists to prevent its annihilation[25], and it could be explained by the attraction of two Bloch points with opposite topological charges, in analogy to the nature of electric charges. However, it should be noted that in micromagnetics the total energy of a Bloch-point system is mesh-dependent, because the energy around the Bloch points depends on the selection of the mesh size for the simulations[14,18,19]. For example, our previous work showed that the required magnetic field to collapse a single skyrmion increases exponentially with linearly decreasing of the mesh size[18]. On the other hand, the barrier heights can also be affected by many factors, such as the choice of materials, film thickness, and applied field strength, which may lead to variations even in several orders of magnitude[6,38]. Therefore, in this work we focus mainly on the comparisons of the proposed transitions paths regarding the relative barrier heights, which should be treated as estimations rather than exact magnitudes.

The Bloch-point propagation alters the "global topological charge" $Q_g$ [Eq. (2)], and could provide us with a different perspective during transitions from the topologically trivial homochiral state to a nontrivial spin texture. In our cases, the $Q_g$ of the HC state features a non-zero value because the spins near the lateral edges tend to twist in-plane by magnetostatic interactions. Then as the Bloch points are created, their propagation through the film induces an accumulation of $Q_g$, accompanied by the increasingly pronounced $B_z^e$ field. In our three-dimensional systems, the $Q_g$ evolutions follow relatively smooth paths. They are in resemblance to the second-order transitions induced by monopole-like topological point defects (Bloch points). The mechanism is different from the radially symmetrical collapse of a single skyrmion[39], where the topological charge is modulated in a steep manner with the absence of a topological point defect. Interestingly, such mechanisms can be analogous to the radical/gradual melting of the topological charge in skyrmion lattices, where the transition modes are characterised as first- or second-order transitions due to the mediation of a high $\mathcal{K}_u$ defect[51].

Moreover, during the transitions we find the magnitude of the topological Hall resistivity $\rho_{xy}^{TH}$ is approximately proportional to the "global topological charge" $Q_g$. In order to elaborate their intrinsic connection, we start with the two-dimensional form of the $\rho_{xy}^{TH}$:

$$\rho_{xy}^{TH} = PR_0\phi_0 \mathcal{Q}_{2D}, \quad (3)$$

with the spin polarisation $P$, effective density of charge carriers $R_0$, and the flux quantum $\phi_0$[8,33,34,52]. $\mathcal{Q}_{2D}$ here indicates the average topological charge density in the area $A$ and can be expressed as $\mathcal{Q}_{2D} = Q_{2D}/A$. This formula has been extended as $\mathcal{Q} = \frac{1}{N}\sum_n Q_{2D,n}/A$, which is able to apply on magnetic multilayer structures with the $Q_{2D,n}$ at $n^{th}$ ferromagnetic layer[35]. As a multilayer is effectively a stack of multiple quasi-two-dimensional systems, we could proceed to a continuous form $\mathcal{Q} = \frac{1}{LA}\int Q_{2D}(z)\,dz$ by referring to the effective thickness of the ferromagnetic layer $L$. It can be further simplified as the average density of the "global topological charge" $Q_g$ for a ferromagnetic system with volume $V$, and therefore, leads to a concise expression:

$$\rho_{xy}^{TH} = PR_0\phi_0 \frac{Q_g}{V}. \quad (4)$$

Then its variation attributed to the Bloch-point propagation is

$$\Delta\rho_{xy}^{TH} = PR_0\phi_0 \frac{\Delta Q_g}{V} = PR_0\phi_0 \frac{\Delta z_{BP}}{V}. \quad (5)$$

It suggests a general correlation between the topological Hall transport property and topological charge even in a bulk system with "considerable" sizes in all three dimensions, in which the nontrivial spin textures usually possess complex spin orientations. This finding is in excellent agreement with the theoretical manifestation of the topological Hall conductivity $\sigma_{xy}^{\text{TH}}$ by Redies et al.[22], where the $\sigma_{xy}^{\text{TH}}$ of chiral bobbers grows linearly as a function of the film thickness $L$ and the $\sigma_{xy}^{\text{TH}}$ of skyrmion tubes is approximately independent of $L$.

**Tailoring energy barrier by modulating surface properties.**
In previous sections, we discuss the energy barriers of the transitions among different textures. Here we propose that modifying the surface properties, such as engineering the additional DMI in the magnetic system, will enable the modulations on the magnetisation profile and thermostability of spin textures. In theory, an interfacial type of DMI arises from the inversion symmetry breaking at the interface[12]; in reality, such an energy term can be introduced by the existence of mechanical strains originating from the magnetoelastic interactions near the chiral magnet/substrate interfaces[9,41,53–55]. Recent studies also provide various novel methods to manipulate mechanical strains, including utilising different thermal strain properties (thermal expansion coefficient) of the chiral magnet and substrate[54], and controlling the ferroelectric polarisation of the substrate by electric fields[55,56]. It should be noted that such modulation may also involve the variation of magnetocrystalline anisotropy[9,44], whose contribution is not considered in this work but definitely deserves further studies.

To understand the effect of the additional symmetry breaking on the stability of spin textures, here we develop a 2D hybrid-DMI model with a mixture of the interfacial DMI $\mathcal{D}^{\text{int}}$ and bulk DMI $\mathcal{D}^{\text{bulk}}$ based on previous theoretical works[1,12,18]. The energy density of the system is written as:

$$\mathcal{E} = \mathcal{E}_{\text{exc}}(\mathbf{m}) + \mathcal{E}_{\text{DMI}}(\mathbf{m}) - M_s \mathbf{B}_{\text{ext}} \cdot \mathbf{m}, \tag{6}$$

including the contributions of (1) exchange energy $\mathcal{E}_{\text{exc}}$, (2) DMI energy $\mathcal{E}_{\text{DMI}}$ and (3) Zeeman energy. The spherical coordinate is introduced to describe the magnetisation $\mathbf{m} = (\sin\theta\cos\phi, \sin\theta\sin\phi, \cos\theta)$ and the polar coordinate is presented for the space $\mathbf{r} = (r\cos\psi, r\sin\psi)$. When a rotationally symmetric profile is considered with solutions of the out-of-plane component $\theta = \theta(r)$ and the in-plane component $\phi = \phi(\psi)$, the exchange energy density $\mathcal{E}_{\text{exc}}$ can be reduced to

$$\mathcal{E}_{\text{exc}}(\theta,\phi,r,\psi) = \mathcal{A}\left[\theta_r^2 + \frac{1}{r^2}\sin^2\theta\,\phi_\psi\right], \tag{7}$$

where $\mathcal{A}$ is the exchange stiffness, and the subscripts indicate partial derivatives. Chiral magnets usually have the $D_n$ symmetry due to their non-centrosymmetric crystal structure. Thus, the bulk DMI $\mathcal{D}^{\text{bulk}}$ is denoted with the energy density $\mathcal{E}_{\text{DMI}}^{\text{bulk}}$ in the form of[1,47,57]:

$$\mathcal{E}_{\text{DMI}}^{\text{bulk}}(\theta,\phi,r,\psi) = \mathcal{D}^{\text{bulk}}\sin(\phi-\psi)\left[\theta_r + \frac{1}{r}\sin\theta\cos\theta\,\phi_\psi\right]; \tag{8}$$

on the other hand, when the structural inversion symmetry is broken, an interfacial DMI arises with the energy density $\mathcal{E}_{\text{DMI}}^{\text{int}}$:

$$\mathcal{E}_{\text{DMI}}^{\text{int}}(\theta,\phi,r,\psi) = \mathcal{D}^{\text{int}}\cos(\phi-\psi)\left[\theta_r + \frac{1}{r}\sin\theta\cos\theta\,\phi_\psi\right]. \tag{9}$$

Therefore, for a 2D skyrmion-like texture with chirality $\phi_1 = \phi - \psi$ (Fig. 4), the analytical expression of the total energy density can be rewritten as:

$$\mathcal{E}(\theta,r,\phi_1) = \mathcal{A}\left[\theta_r^2 + \frac{1}{r^2}\sin^2\theta\right] + \mathcal{D}^{\text{bulk}}(\mathcal{D}^{\text{ratio}}\cos\phi_1 + \sin\phi_1)\left[\theta_r + \frac{1}{r}\sin\theta\cos\theta\right] - M_s B_{\text{ext}}\cos\theta, \tag{10}$$

with the DMI ratio $\mathcal{D}^{\text{ratio}} = \mathcal{D}^{\text{int}}/\mathcal{D}^{\text{bulk}}$. It is worth noting that Eq. (10) suggests the densities of exchange energy (first term) and Zeeman energy (third term) are solely dependent on the out-of-plane profile $\theta(r)$ under a given external field $B_{\text{ext}}$. However, the DMI energy density (second term) can also be altered by in-plane chirality $\phi_1$, which corresponds to an introduction of $\mathcal{D}^{\text{int}}$. The calculated landscape of the total energy density of a two-dimensional skyrmion is shown in the background colour of Fig. 5a, as the functions of the DMI ratio $\mathcal{D}^{\text{ratio}}$ and in-plane chirality $\phi_1$. It is noted that because $\mathcal{D}^{\text{bulk}}$ denotes positive values in this work, the energetically preferred $\phi_1$ mainly lies in the range of $\phi_1 < 0$. Then in the following, the $\phi_1$ is represented by its magnitude (absolute value $|\phi_1|$). As a result, for a 2D skyrmionic magnetisation profile, as the $\mathcal{D}^{\text{ratio}}$ varies from "-1" ($\mathcal{D}^{\text{int}} = -\mathcal{D}^{\text{bulk}}$) to "+1" ($\mathcal{D}^{\text{int}} = \mathcal{D}^{\text{bulk}}$), the energetically preferred chirality is changed $1/4\,\pi$ to $3/4\,\pi$ (dots in Fig. 5a).

Then we illustrate the in-plane chirality of the bottom-surface profiles of bCB (triangles) and SkT (squares), extracted from numerical simulations at $B_{\text{ext}} = 0.97 B_0'$ (Fig. 5a). Although the system is purely dominated by the bulk DMI ($\mathcal{D}^{\text{ratio}} = 0$), their chiralities are smaller than that of a Bloch-type skyrmion ($1/2\,\pi$). This can be attributed to the effect of the "surface twist" due to the geometrical confinement[1,17] and the non-uniform distribution of the magnetostatic energy through the film[18,46]. With the presence of the interfacial DMI near the bottom surface, a negative $\mathcal{D}^{\text{int}}$ ($\mathcal{D}^{\text{ratio}} < 0$) tends to draw the bottom-surface chiralities close to the analytical energy minimum, whereas a positive $\mathcal{D}^{\text{int}}$ ($\mathcal{D}^{\text{ratio}} > 0$) plays an opposite role and "pushes" the chiralities away from the analytical energy minimum. Such $\mathcal{D}^{\text{int}}$-induced modulation on the total energy of topological spin textures is confirmed by the numerical results (Fig. 5b–c). Because the magnetisation profiles of the HC state, tCB, and DS are composed of similar conical phases near the bottom surface, their relative total energies are approximately independent of $\mathcal{D}^{\text{int}}$. However, the $\mathcal{D}^{\text{int}}$ modulation has a more pronounced effect on the textures whose bottom-surface profiles are nontrivial, such as bCB, CB pair, and SkT. In particular, we also observe a high positive $\mathcal{D}^{\text{int}}$ ($\mathcal{D}^{\text{ratio}} > 0.8$ in this work) can even trigger a partial annihilation of a CB pair to the tCB state (dashed purple line in Fig. 5b–c).

A parametric study is conducted to investigate the space of ground-state textures (Fig. 6). In the pure bulk DMI ($\mathcal{D}^{\text{int}} = 0$) regime we have already known that as the magnetic field $B_{\text{ext}}$ increases, the HC (helical state), SkT, or HC (conical/saturated state) is energetically preferred respectively[4,21,25] (Fig. 1g). The presence of a negative $\mathcal{D}^{\text{int}}$ enhances the stability of bottom-surface textures and thus enlarges the ground-state region of skyrmion tubes. Especially at the high magnetic field ($B_{\text{ext}} = 0.97 B_0'$), a ground-state bCB is illustrated with $\mathcal{D}^{\text{ratio}} \leq 0.6$ in this work (Fig. 5b–c), and a bCB "pocket" exists in a triangular region (blue down triangles in Fig. 6) between the regions of the skyrmion tube and homochiral state. On the other hand, a positive $\mathcal{D}^{\text{int}}$ tends to destabilise the local textures, which yields a broad ground-state region of tCBs.

To elucidate the effect of $\mathcal{D}^{\text{int}}$ modulation on the preference of transition paths, especially towards the ground-state textures, we calculate the variation of the minimum energy paths along the HC – CB(s) – SkT at $B_{\text{ext}} = 0.97 B_0'$ (Fig. 7a–d). Reaction coordinates are scaled for the convenience of comparing the energies of local minima and saddle points in different paths. As a result, the transition paths can be modulated by $\mathcal{D}^{\text{int}}$, which is highly related to the change of bottom-surface chiralities. Here we choose the texture growth direction (HC → SkT) as examples. During the transitions (type I) mediated by intermediate single CBs, Fig. 7a shows the minimum energy paths are independent of $\mathcal{D}^{\text{int}}$ before reaching the saddle points between tCB and SkT, where the Bloch point is created and propagates to the bottom surface. After the Bloch point unfolds to a 2D skyrmion, the bottom-surface chirality will be relaxed in terms of $\mathcal{D}^{\text{int}}$, and the total energy of the entire texture becomes $\mathcal{D}^{\text{int}}$-dependent. Similarly, in Fig. 7b the variation of minimum energy paths mainly occurs in the initial stage where single bCBs are formed from the HC state.

Moreover, we also extract the energy barriers of these transitions during the texture growth from HC to SkT (Fig. 7e) and texture degradation from SkT to HC (Fig. 7f). The increase/decrease of barrier heights are closely related to the energy variations of the textures, and more specifically, of the local chirality near the bottom surface. Modulated by a positive $\mathcal{D}^{\text{int}}$, the relative total energies of bCB, CB pair and SkT are raised (Fig. 5b), in correspondence with the increase of texture-growth barriers and decrease of texture-degradation barriers. It is intuitive that in the $\mathcal{D}^{\text{ratio}} = 0$ regime, the formations of tCB and bCB from the HC state are energetically identical. Moreover, as the nanodisk is relatively thick in this work, the formation/annihilation of a single CB near the top (bottom) surface has a minor effect on the transition occuring near the bottom (top) side, e.g. referring to the energy barriers of a tCB growth in the processes HC → tCB and bCB → CB pair (Fig. 7e1 and e5], as well as those of a bCB growth in the processes HC → bCB and tCB → CB pair (Fig. 7e2 and e3). However, the introduction of $\mathcal{D}^{\text{ratio}}$ is able to deterministically increase/decrease the probability of a transition by locally lowering/raising its energy barrier, and thus achieve an artificial control on the transitions among these topological spin textures. Especially, the local energy minimum of intermediate CB pairs (Fig. 7c) can be smoothed out by lowering the energy barriers close to zero (Fig. 7f3). Therefore, the type-II transition can be suppressed by destabilising the CB pair during the transitions. It suggests that the ground state can be tailored by the introduction of interfacial DMI near the surface of chiral magnets, and provides an effective transition route in addition to the conventional field-induced or thermal-induced transitions[3,4,18,38,45,58].

Among these topological spin textures, the stabilisation of metastable states is more generic, especially for chiral bobbers and dipole strings in the whole range of $B_{\text{ext}}$ (Fig. 1i). The occurrence of their thermal transitions needs to surpass finite energy barriers, and extra excitations (such as thermal effect) are necessary for escaping from the local energy minima. Therefore, future applications that exploit metastable spin textures are worth considering, and recent work has contributed to the idea by the experimental observation of metastable chiral bobbers in a wide range of temperature and external magnetic field[4]. In general, the stability can be quantified by a lifetime $\tau$, which is relevant to the barrier height $\Delta E$ and temperature $T$. It follows the Arrhenius expression: $\tau = \tau_0 \exp(\Delta E/k_\mathrm{B} T)$, where $k_\mathrm{B}$ is the Boltzmann constant, and $\upsilon = \tau_0^{-1}$ indicates the rate prefactor (also interpreted as "attempt frequency" in literature)[59–61]. Experimentally, the barrier height is usually assumed to be linearly dependent on temperature, originating from the proportional relationship between the temperature and square of the local magnetic moment[62,63]. For thermally activated magnetisation reversals in nanostructures, the value of the "attempt frequency" was estimated in the order of GHz[39,64,65]. On the aspect of the information storage technologies, a ten-year lifetime usually requires $\Delta E/k_\mathrm{B}T > 60$ at room temperature ($T = 300\,\mathrm{K}$)[66]. However, recent studies have also shown that the Arrhenius prefactor can be greatly underestimated without considering the entropy change during the transitions[5,37,40]. On the other hand, in this work, we focus on the transitions in a rotationally symmetric manner, where the processes mainly occur in the centre of the nanodisk, whereas other transition paths with different barrier heights may also exist. For example, Cortés-Ortuño et al. showed that the transitions between HC state and skyrmion via geometric boundaries have much lower energy barriers compared to those Bloch-point-mediated mechanisms[39]. Therefore, our discussion of this section does not intend to provide an all-inclusive solution such as quantitatively altering the lifetime of a topological spin texture, but aims to evaluate the feasibility of exploiting the surface-property modulation for transitions among various textures in chiral magnets.

## Conclusion

We have demonstrated the thermal stabilities and transitions of various topological spin textures in chiral magnetic nanodisks, including those having been predicted or observed so far: single chiral bobbers, chiral-bobber pairs, dipole strings, and skyrmion tubes. All these textures can be stabilised as metastable states in a wide range of external magnetic fields, and critically, skyrmion tubes are the ground state of the system within specific fields. Different textures can be classified by their topological Hall signals (exemplified by topological Hall resistivity $\rho_{xy}^\mathrm{TH}$ in this work) due to their nontrivial topological properties. Its magnitude is proportional to the density of the "global topological charge", defined as the volume integral of the emergent magnetic field.

To identify the thermal transitions among these spin textures, we investigate three main minimum energy paths between the homochiral state and skyrmion tube, via the intermediate presence of single chiral bobbers, chiral-bobber pairs, or dipole strings. All the minimum energy paths converge to the mechanisms mediated by Bloch-point singularities. During the transitions, the topological protection property is manifested as finite energy barriers, and the saddle points are associated with the creation or destruction of Bloch points.

Furthermore, on the basis of our analytical model, we propose a scenario for modifying the energy of surface states by engineering the ferromagnetic/nonmagnetic interface properties, such as introducing additional interfacial DMI. This directly modulates the energy barriers of transitions, and could even facilitate the emergence of ground-state nontrivial textures (such as chiral bobbers) in broad parameter spaces of the magnetic field and interfacial DMI. The proposed surface modulation method could act as an effective technique for selectively switching local spin textures regarding their depth-resolved magnetisation profile, in addition to applying magnetic field or heat bath.

This work strives to lay a stone in the fundamental research on complex topological spin textures, which have a great potential of being utilised on three-dimensional spintronic devices, equipped with multi-bit functionalities for data encoding, neuromorphic computing, and interconnect devices[67,68]. As prerequisites of device-level implementations, future studies focusing on pertinent topics could be particularly intriguing, such as (1) in-depth investigations on the metastability of topological spin textures, including theoretical demonstrations of the lifetime relying on the harmonic transition-state theory (HTST)[61] or forward flux sampling (FFS) method[40]; (2) techniques for decisive control of the stability via directly engineering magnetic properties[35,55] or even local defects[18,69]; (3) inspired by the skyrmion Hall effect, the current-driven dynamics of three-dimensional textures may incorporate rich dynamics with distinct Hall angles[7,26,29]. These results offer greater possibilities for future experimental demonstrations, which are now increasingly within reach.

## Methods

**Micromagnetic simulation.** Topological spin textures in this work are simulated in B20-type compound FeGe[2–5,36,45,53]. In the family of chiral magnets, FeGe is a promising candidate for potential applications concerning its comparatively high Curie point near room temperature[3,45,53] and pronounced topological Hall signal[3,53]. Because of the radially symmetric profiles of these topological spin textures, the geometry of the magnetic system is considered as a nanodisk. Then the diameter is set to $4L_D$ to minimise the confinement effect from lateral boundaries; the thickness $2L_D$ is above a critical value ($0.68L_D$ in Ref. [43]) so that the conical state (Fig. 1b) will not be completely suppressed by the effect of surface-induce twists[17,43]. Here $L_D = 4\pi\mathcal{A}/|\mathcal{D}| \approx 70$ nm is the helical period of FeGe with exchange stiffness $\mathcal{A} = 8.78$ pJ m$^{-1}$ and DMI constant $\mathcal{D} = 1.58$ mJ m$^{-2}$ in Ref.[36]. Nanodisks are discretised into 2 nm × 2 nm × 2 nm cubic cells, whose length is smaller than half of the exchange length $\delta_{ex}/2 = \sqrt{2\mathcal{A}/\mu_0 M_s^2}/2 \approx 4.9$ nm, with the vacuum permeability constant $\mu_0$ and the saturation magnetisation $M_s = 384$ kA m$^{-1}$ in Ref. 36. Our discussions involve the micromagnetic expression of the total energy density:

$$\mathcal{E} = \mathcal{A}(\nabla \mathbf{m})^2 + \mathcal{D}\mathbf{m}(\nabla \times \mathbf{m}) - M_s \mathbf{B}_{ext} \cdot \mathbf{m} - \frac{1}{2} M_s \mathbf{B}_{demag}(\mathbf{m}) \cdot \mathbf{m}, \qquad (11)$$

including the contributions of (1) exchange interaction, (2) isotropic bulk DMI for the $D_n$ symmetry class, (3) Zeeman coupling with the external magnetic field $\mathbf{B}_{ext}$, and (4) magnetostatic interaction in the demagnetising field $\mathbf{B}_{demag}(\mathbf{m})$. $\mathbf{m}$ is the vector field of the normalised magnetisation with constant magnitude $M_s$. Micromagnetic simulations are performed using Fidimag[70] for calculations of equilibrium states of topological spin textures, where a magnetic configuration is relaxed based on the Landau-Lifshitz equation of motion [Eq. (11)][71,72], followed with the minimisation of the total energy by the steepest descent method[73]. The geodesic nudged elastic band (GNEB) method is used to calculate minimum energy paths between the proposed equilibrium states, so that the energy barriers along transitions can be quantified[74].

**Electronic transport property.** The calculations of the electronic transport properties are implemented by Kwant[75] based on simple cubic scattering lattices, with four ferromagnetic leads attached to each nanodisk (Fig. 2a). The topological Hall resistivity is calculated based on the multi-probe Landauer-Büttiker formalism[76] by analogy with the work in Refs.[31,32,35]. A reading current is applied from the left lead to the right one with a negligibly small magnitude, thus the current-induced motion of spin textures[13] can be excluded. The Hall voltage attributed to the deflection of electrons into the transverse direction is measured by the voltage drop from the top lead to the bottom one. Then the interaction of electrons with local spin textures is described by a single-orbital tight-binding model with the Hamiltonian:

$$\mathcal{H}_e = -t \sum_{\langle i,j \rangle} (c_i^\dagger c_j + \text{H.c.}) - J_\text{H} \sum_i c_i^\dagger \mathbf{m}_i \cdot \sigma c_i, \tag{12}$$

where $t$ indicates the hopping integral between nearest-neighbour sites, $c_i^\dagger$ ($c_i$) is the creation (annihilation) operator, "H.c." is the abbreviation of the Hermitian conjugate of the other term, and $J_\text{H}$ is the exchange energy of the coupling between the spin of conduction electrons $\sigma$ and the local magnetisation $\mathbf{m}_i$.

## Data availability
All data that support the findings of the paper are available from the corresponding author upon reasonable request.

## Acknowledgements

Financial support by the EP/V028189/1 grant is gratefully acknowledged. The authors would like to acknowledge the assistance given by Research IT and the use of the Computational Shared Facility at The University of Manchester. Y.L. and R. C. acknowledge the funding support by the University of Manchester and the China Scholarship Council.

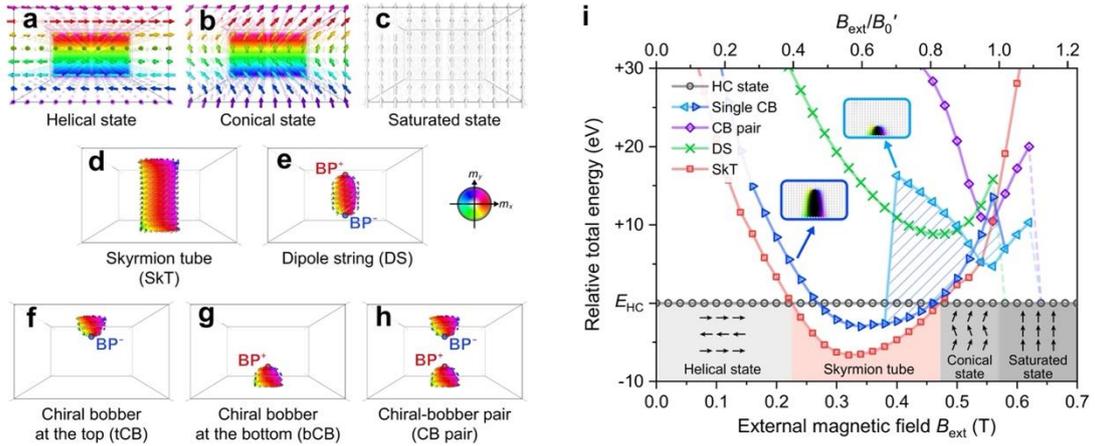

**Fig. 1 Various topological spin textures in chiral magnets**. **a-c** Homochiral state, including **a** helical state at zero magnetic field with the wave vector (propagation direction) perpendicular to the film surface, **b** conical state with the wave vector perpendicular to the field direction, and **c** saturated state with all magnetisation pointing along the field direction. **d** Skyrmion tube with spatially non-uniform chiralities along the tube. **e** Dipole string with two Bloch points at the upper and lower ends. **f-h** Chiral bobber states, including a single chiral bobber situated near the **f** top or **g** bottom surface, and **h** chiral-bobber pair near both surfaces. Two types of Bloch points with opposite topological charges (BP$^+$ and BP$^-$) are located at the ends of dipole strings and chiral bobbers. The colour palette represents the magnetisation direction, and isosurfaces indicate the positions of $m_z = 0$. **i** Variation of the total energy of these textures as a function of the magnetic field. The total energies are plotted with reference to the homochiral state ($E_{HC}$) and the background colours indicate the ground states. Especially, the single chiral bobber states include two metastable types with the penetration depth $L_P < L_D$ (cyan left triangles ◁) or $L_P > L_D$ (blue right triangles ▷), and can be switched from one to another with a hysteretic behaviour. Solid lines connecting the data points are guides to the eye.

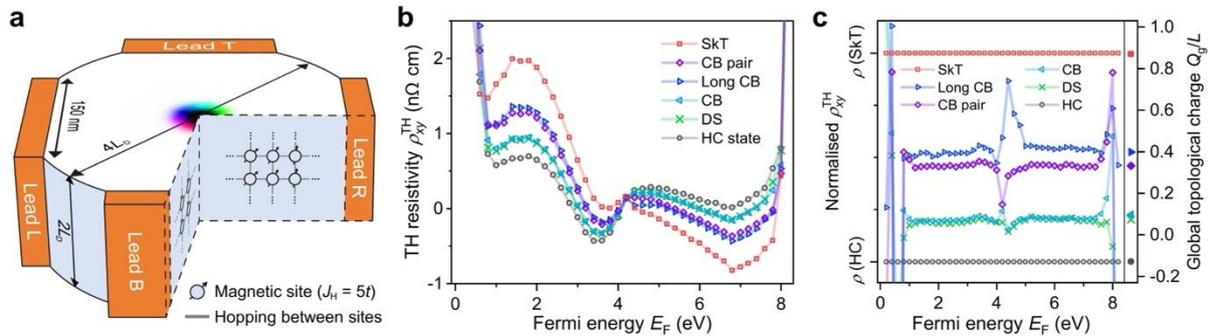

**Fig. 2 Topological Hall transport properties of various topological spin textures**. **a** Schematic of the four-terminal setup, with a single topological spin texture placed in the disk centre. **b** Topological Hall resistivity $\rho_{xy}^{TH}$ of topological spin textures as a function of Fermi energy $E_F$, and only $E_F = 0$ is visualised. Solid lines connecting the data points are guides to the eye. **c** Normalised topological Hall resistivities $\rho_{xy}^{TH}$ by rescaling the values with respect to those of the HC state and skyrmion tube at each $E_F$. As a comparison, the global topological charge $Q_g$ is presented as dots in the units of film thickness $L = 2L_D$.

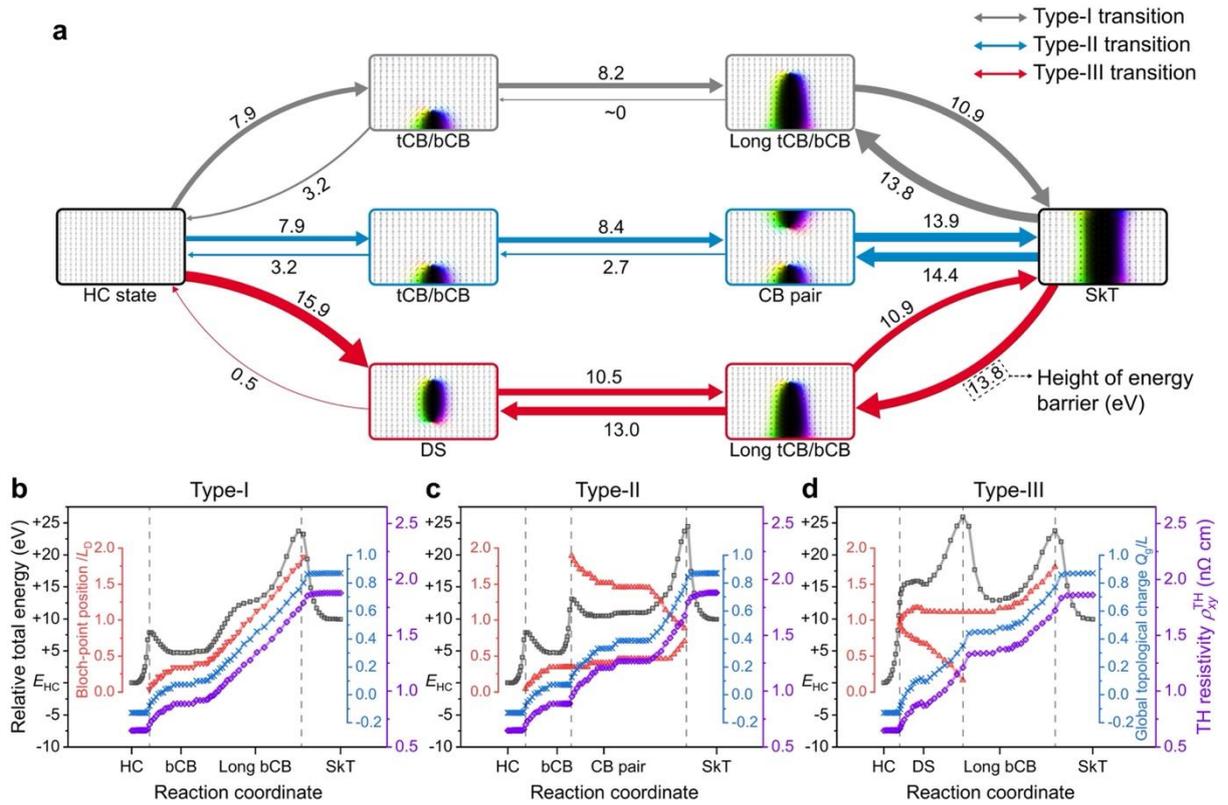

**Fig. 3 Three types of transition paths between homochiral state and skyrmion tube**. **a** Minimum energy paths and the associated energy barriers. Three possible transition paths are proposed, which are characterised by the intermediate presence of single chiral bobber (type I), chiral-bobber pair (type II) and dipole string (type III). **b-d** The minimum energy paths (black squares □) of transitions, as well as the associated Bloch-point position (red rectangles △▽), global topological charge $Q_g$ (blue crosses ✕), and topological Hall resistivity $\rho_{xy}^{TH}$ (purple diamonds ◇). $Q_g$ is illustrated in the units of thickness $L = 2L_D$ and $\rho_{xy}^{TH}$ is calculated at $E_F = 2t$. Dashed lines indicate the creation or destruction of Bloch point(s). The reaction coordinates are scaled for the convenience of comparing the saddle points. Solid lines connecting the data points are guides to the eye.

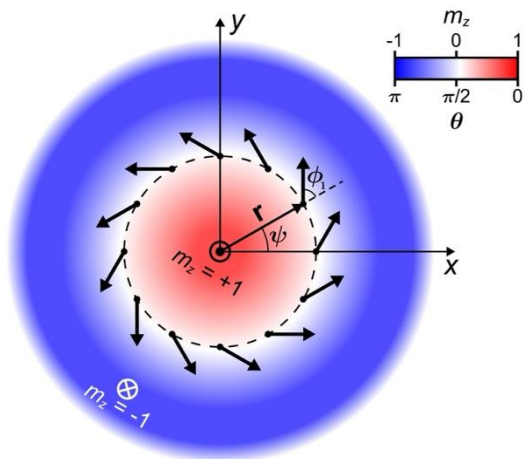

**Fig. 4 Top-view schematic of the magnetisation profile of a magnetic skyrmion in a two-dimensional plane**. The skyrmion has an axisymmetric profile, with the magnetisation in the core pointing to the $m_z = +1$ direction (polarity $p = 1$) whereas in the surrounding areas pointing to the $m_z = -1$ direction. The in-plane orientation of the magnetisation is represented as thick arrows, and the skyrmion chirality $\phi_1$ is defined as the angle between the in-plane magnetisation $\phi$ and its position vector $\psi(r)$.

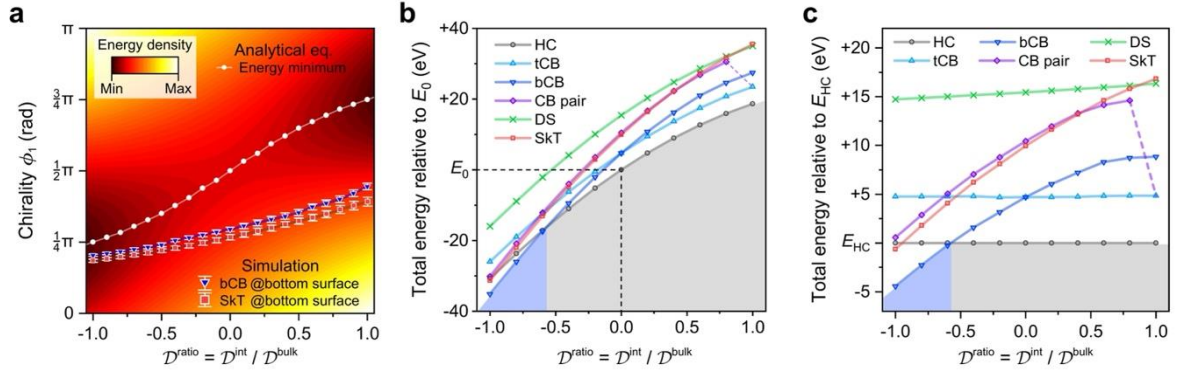

**Fig. 5 Stabilities of topological spin textures as a function of the bottom-surface $\mathcal{D}^{\mathrm{int}}$. a** Total energy density (background colour) of a 2D skyrmion with respect to the DMI ratio $\mathcal{D}^{\mathrm{ratio}} = \mathcal{D}^{\mathrm{int}}/\mathcal{D}^{\mathrm{bulk}}$ and its in-plane chirality $\phi_1$, calculated by the analytical model [Eq. (10)]. The local energy minimum at each $\mathcal{D}^{\mathrm{ratio}}$ is labelled by dots •. The bottom-surface chiralities $\phi_1$ of bCB (blue down rectangles ▽) and SkT (red squares □) are extracted from numerical simulations at $B_{\mathrm{ext}} = 0.97 B_0'$. Error bars indicate the minor variation of chiralities in space. The chiralities are represented by their absolute values $|\phi_1|$. **b** Evolution of the total energy as a function of DMI ratio $\mathcal{D}^{\mathrm{ratio}}$ with reference to $E_0$, the total energy of the homochiral state at $\mathcal{D}^{\mathrm{ratio}} = 0$. **c** The total energy of homochiral state $E_{\mathrm{HC}}$ and topological spin textures at different $\mathcal{D}^{\mathrm{ratio}} = 0$. The dashed purple lines indicate the partial annihilation of a CB pair to a tCB at high $\mathcal{D}^{\mathrm{ratio}}$. The background colour illustrates the ground states, and solid lines connecting the data points are guides to the eye.

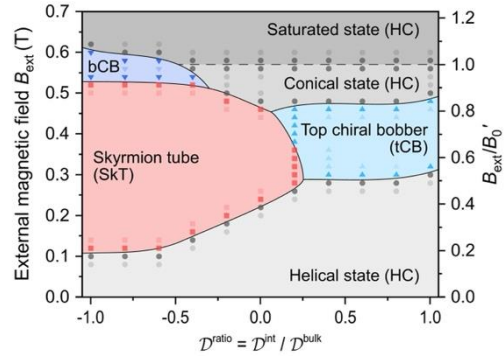

**Fig. 6 Phase diagram of ground states as functions of the DMI ratio $\mathcal{D}^{\mathrm{ratio}}$ and external magnetic field $B_{\mathrm{ext}}$.** The diagram is plotted without considering the effect of temperature ($T = 0$ K). The dashed line indicates the second-order transition between the conical and saturated state.

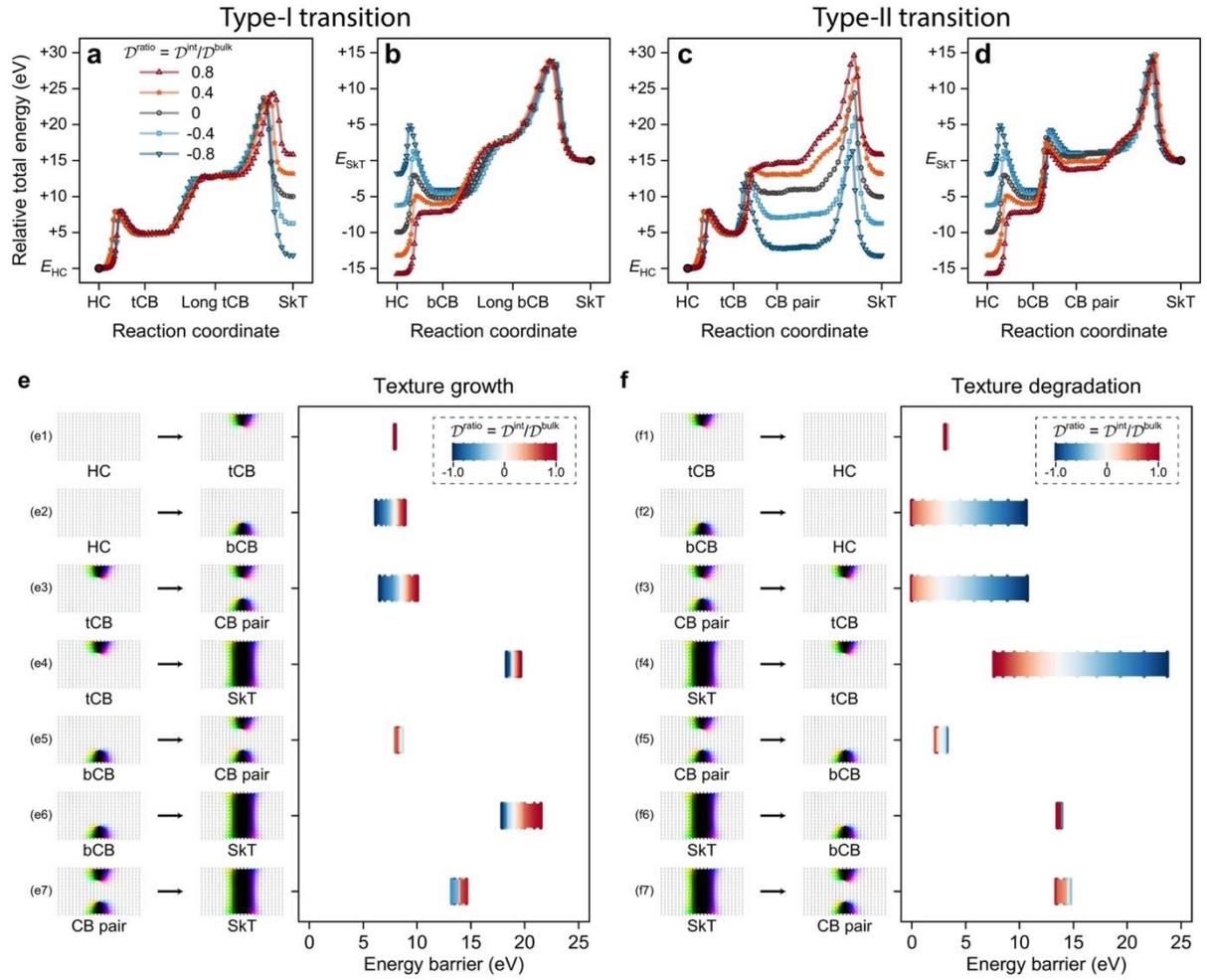

**Fig. 7 Modulation on the transitions and energy barriers by DMI ratio $\mathcal{D}^{\text{ratio}}$ at the bottom surface**. **a-d** Minimum-energy paths via the intermediate **a** tCB, **b** bCB, **c** tCB + CB pair, and **d** bCB + CB pair. The total energy of images along the paths are plotted with reference to the HC state or SkT, and the image distances in reaction coordinates are scaled for the convenience of comparing the energies of local minima and saddle points. **e-f** Variation of energy barriers of **e** texture growth and **f** texture degradation as a function of $\mathcal{D}^{\text{ratio}}$. The texture configurations of the transitions and associated barrier heights are illustrated in the left and right panels, respectively. The value of $\mathcal{D}^{\text{ratio}}$ is indicated by the colour palette.

# Supplementary information:
# Tailoring energy barriers of Bloch-point-mediated transitions between topological spin textures

Yu Li[1,2*], Yuzhe Zang[1], Runze Chen[1], Christoforos Moutafis[1*]

[1]Nano Engineering and Spintronic Technologies (NEST) Group, Department of Computer Science, University of Manchester, Manchester M13 9PL, United Kingdom.
[2]Frontier Institute of Chip and System, Fudan University, Shanghai, 200438, China.
*Email: YuLi.nano@outlook.com; Christoforos.Moutafis@manchester.ac.uk.

## Supplementary note 1: Geometry-dependent critical field of conical-saturated transition

In chiral magnets, the critical field $B_c$ could indicate the transition field between the conical and saturated states. They will undergo the second-order transitions with continuous variations of the magnetic susceptibility $\chi = \partial \mathbf{m}/\partial \mathbf{B}_{\text{ext}}$ (second derivative of the energy) as a function of the external magnetic field $\mathbf{B}_{\text{ext}}$. Its analytical solution can be expressed as: $B_0 = \mathcal{D}^2/2\mathcal{A}M_s$ ($\sim 0.37$ T for FeGe) regarding the micromagnetic parameters[S1–S3]. Its derivation is based on the system without the contribution of magnetostatic interactions. However, we find the magnetostatic interaction could play a significant role in modulating conical-saturated transitions, which is dependent on the sample geometry (Fig. S1). It gives rise to a shape anisotropy which has a tendency of reducing the total magnetic moment by driving the magnetisation vectors away from the surface normal. Therefore, in a relatively flat nanodisk with a small length-to-diameter ratio, the magnetisation tends to align in the $xy$ plane accompanied by a high critical field ($B_\text{c} > B_0$), in analogy to a hard-axis model with a negative uniaxial anisotropy constant ($\mathcal{K}_{\text{u}1} < 0$); in contrast, for a large length-to-diameter ratio, the system can be intuitively regarded as a nanorod, where an easy axis points along the rod ($z$ axis) with a positive $\mathcal{K}_{\text{u}1}$, which has an effect of lowering the critical field ($B_\text{c} < B_0$).

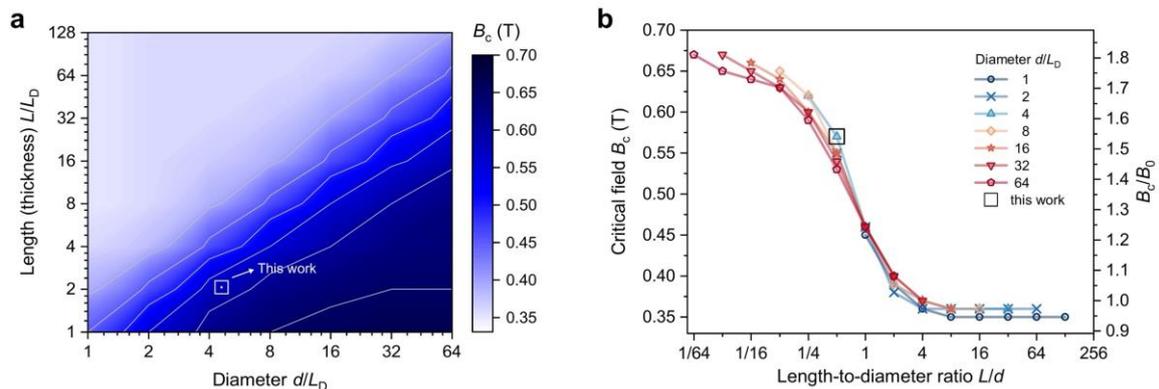

**Fig. S1 Critical field of transitions between the conical and saturated states. a** Contour plot of the critical field $B_c$ (background colour) regarding the sample diameter $d$ and length (thickness) $L$, in the units of $L_D = 70$ nm. **b** Critical field $B_c$ shown as a function of the length divided by diameter $L/d$. The nanodisk geometries simulated in this work are indicated by the square, with the diameter of $4L_D$ and the length (thickness) of $2L_D$, and the adjusted critical field is $B'_0 = 0.57$ T. $B_0 = 0.37$ T indicates the critical field of FeGe chiral magnets without considering the effect of magnetostatic interactions.

In this work, the geometry of nanodisks is chosen with the diameter $d = 4L_\text{D}$ and length (thickness) $L = 2L_\text{D}$ unless specified elsewhere, and it leads to an adjusted critical field $B_0' = 0.57\,\text{T}$. The results are directly extracted from the micromagnetic simulations, and the critical fields with respect to the sample geometry are shown in Fig. S1.

## Supplementary note 2: Metastable chiral bobbers in $4L_\text{D}$-thick nanodisks

We have shown that single chiral bobbers can exist with two different penetration depths $L_\text{P}$ in $2L_\text{D}$-thick nanodisks (Fig. 1i in the main text). The presence of finite energy barriers between them prevents their transitions from one type to another (Fig. 3b in the main text). These objects can be imaged by various techniques, such as Lorentz transmission electron microscopy (LTEM)[S4–S6], TEM mode of off-axis electron holography[S5–S8], scanning transmission x-ray microscopy (STXM)[S4], and x-ray photoemission electron microscopy (XPEEM)[S9]. As chiral bobbers with different $L_\text{P}$ have unique topological Hall responses (Fig. 2b in the main text), their detection and distinction can also be realised by electrical measurements. It paves the way for being utilised as multi-bit information carriers, and theoretically, the total number of possible states increases as a function of the thickness (in the order of $nL_\text{D}$). Here we present four types of chiral bobbers that can be stabilised as metastable states, at the bottom surface (bCB) of $4L_\text{D}$-thick nanodisks (Fig. S2). HC states and skyrmion tubes can be regarded as special chiral bobbers with no penetration into the film ($L_\text{P} = 0$, Fig. S2a) and penetration through the film ($L_\text{P} = 4L_\text{D}$, Fig. S2f), respectively. Based on the relative total energy (Fig. S2g) and the minimum energy path (Fig. S2h), these chiral bobbers have distinct energy landscapes, and although bCB$_4$ and SkT are approximately identical in energy, a large energy barrier ($> 10\,\text{eV} \,@B_\text{ext} = 0.50\,\text{T}$ in this work) still exists between their transition paths.

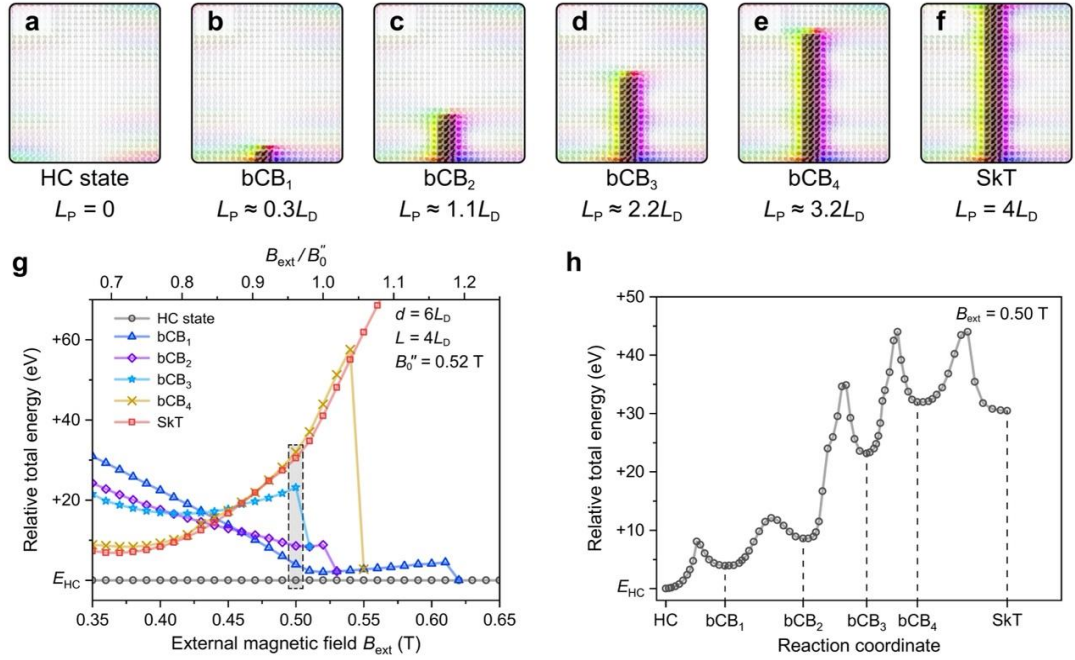

**Fig. S2 Chiral bobbers in $4L_D$-thick nanodisks**. **a-f** A single chiral bobber at the bottom surface (bCB) can be stabilised with different penetration depths $L_P$ (in the units of $L_D = 70$ nm for FeGe). The sample geometry is a nanodisk with the diameter of $d = 6L_D$ and thickness of $L = 4L_D$, and the critical field of conical-saturated transitions is $B_0' = 0.52$ T. **g** Relative total energy of these textures as a function of the magnetic field $B_{ext}$. **h** Minimum energy path between the HC state and SkT, and the local minima correspond to the states at $B_{ext} = 0.50$ T in the dashed square region of **g**.